\documentstyle[12pt,equation,graphicx]{article}
\begin{document}
\newcommand{\be}{\begin{equation}}
\newcommand{\ben}{\begin{subequations}}
\newcommand{\een}{\end{subequations}}
\newcommand{\beq}{\begin{eqalignno}}
\newcommand{\eeq}{\end{eqalignno}}
\newcommand{\ee}{\end{equation}}
\newcommand{\tanb}{\mbox{$\tan \! \beta$}}
\newcommand{\cotb}{\mbox{$\cot \! \beta$}}
\newcommand{\cosb}{\mbox{$\cos \! \beta$}}
\newcommand{\sinb}{\mbox{$\sin \! \beta$}}
\newcommand{\wt}{\widetilde}
\def\gsim{\:\raisebox{-0.5ex}{$\stackrel{\textstyle>}{\sim}$}\:}
\def\lsim{\:\raisebox{-0.5ex}{$\stackrel{\textstyle<}{\sim}$}\:}

\pagestyle{empty}

\begin{flushright}
February 2005\\
\end{flushright}
\vspace*{1.5cm}

\begin{center}
{\Large \bf A Supersymmetric Explanation of the Excess of Higgs--Like Events
  at LEP} \\
\vspace{1cm}
{\large Manuel Drees} \\
\vspace*{6mm}
{\it Physikalisches Institut. d. Universit\"at Bonn, Nussallee 12, 53115 Bonn,
  Germany} 
\end{center}

\begin{abstract}
  Searches for the Standard Model Higgs boson by the four LEP experiments
  found excess events in two mass ranges: a 2.3$\sigma$ excess around 98 GeV,
  and an 1.7$\sigma$ excess around 115 GeV. The latter has been discussed
  widely in the literature, but the former has attracted relatively little
  attention so far. In this paper I explore the possibility of explaining the
  excess near 98 GeV through production of the lighter CP--even Higgs boson in
  the Minimal Supersymmetric Standard Model (MSSM). It is shown that this
  allows to simultaneously explain the excess near 115 GeV through the
  production of the heavier CP--even MSSM Higgs boson. The resulting light
  Higgs sector offers opportunities for charged Higgs boson searches at the
  Tevatron and LHC. Neutral Higgs boson searches at the LHC in the di--muon
  channel are also promising. However, conclusive tests of this scenario may
  have to wait for the construction of a linear $e^+e^-$ collider.

\end{abstract}
\clearpage

\setcounter{page}{1}

\pagestyle{plain}

\section*{1. Introduction}

The search for the Higgs boson of the Standard Model (SM) was one of the
priorities of the four LEP experiments ALEPH, DELPHI, L3 and OPAL. Recently
the final result of this search has been published jointly by these
experiments \cite{leph}. The absence of a clear signal implies that the SM
Higgs boson mass should exceed 114.4 GeV at the 95\% confidence level.

Although no clear signal was found, there are some intriguing hints in these
data. Much has been made \cite{aleph} of the excess of events at the kinematic
edge, for a Higgs mass near 115 GeV. However, this excess comes almost
entirely from a single experiment (ALEPH), and is only visible in a single
final state (with four jets). In contrast, all four experiments see at least a
mild excess near 98 GeV, the ``signal'' being most pronounced in the L3 data
set. Moreover, it exists both in the 4--jet data set and the remaining set of
Higgs candidate events. As a result, while the combination of LEP data
weakened the significance of the excess near 115 GeV to about 1.7 standard
deviations, corresponding to a 9\% background fluctuation probability, the
excess near 98 GeV has a significance of about 2.3 standard deviations, which
corresponds to a background fluctuation probability of only 2\%. This
significance is comparable to that of the deviation of the anomalous magnetic
moment of the muon from its SM prediction \cite{amu}, which has also generated
much interest in recent years.

In this paper I point out that the excess near 98 GeV can easily be
accommodated in the Minimal Supersymmetric extension of the SM (MSSM). Since
the number of excess events corresponds to only about 10\% of the number of
signal events expected for an SM Higgs boson of this mass, the coupling of the
light CP--even MSSM Higgs boson $h$ to the $Z$ boson must be suppressed. This
immediately implies that one must not be in the ``decoupling scenario'', i.e.
the overall mass scale in the MSSM Higgs sector must be quite low. Moreover,
the heavy CP--even Higgs boson $H$ must have couplings quite similar to those
of the SM Higgs boson; it can thus easily be used to explain the excess near
115 GeV. This makes it legitimate to combine the two excesses of events when
assessing the significance with which they favor the MSSM over the SM. In the
limit where the two sets of events are statistically independent, which should
be a good approximation, the total excess amounts to about 3.1 standard
deviations, corresponding to a background fluctuation probability of only
0.2\%. 

This significance as stated applies to situations where the locations of the
excess events are fixed. Since the relevant distribution (in the recontructed
Higgs mass) contains several bins, the probability to find such excesses
somewhere due to statistical fluctuations should be larger than 0.2\%.
However, at least in the framework of the MSSM the total number of bins that
are compatible with the absence of a (clear) signal for $hA$ production is
quite small; note that a small $ZZh$ coupling implies that the $ZhA$ coupling
is nearly maximal. Values of $m_h \lsim 90$ GeV can thus be excluded from the
analysis. Note also that, again in the framework of the MSSM, the $h-H$ mass
difference cannot be arbitrarily small; this further limits the range of $m_h$
where a weak $h$ signal (due to suppressed $hZZ$ coupling) can show up, if the
``signal'' near 115 GeV is interpreted as $ZH$ production. The ``look
elsewhere'' effect should therefore not dilute the total significance by more
than a factor of 5 or so; of course, a reliable estimate of the total
significance of the combined excess would have to come from the experimental
groups.

This scenario can be realized for a wide range of values for the ratio of
vacuum expectation values (VEVs) $\tan\beta$. This means that the light scalar
Higgs $h$, the CP--odd Higgs $A$ and the charged Higgs $H^\pm$, while all
quite light, might couple only weakly to gauge bosons and heavy quarks. As a
result, only the discovery of the heavy scalar $H$ is guaranteed at the LHC in
this scenario. However, since the squared couplings of $H$ differ from those
of the SM Higgs only at the 10\% level, it will be difficult to use $H$
production at the LHC for a decisive test of this scenario. Such a test might
be possible through $t \rightarrow H^+ b$ decays or, at large $\tan\beta$,
through associated $b \bar b h$ and $b \bar b A$ production at the LHC, but
the entire allowed parameter space can probably only be probed at a future
linear $e^+e^-$ collider.

The possibility that the excess near 115 GeV could be due to $ZH$ production
is also discussed in \cite{weiglein,intense}, and, for a CP--violating
scenario, in \cite{carena0}; however these papers do not mention the by now
more significant excess of events near 98 GeV. A scenario with three
(relatively) light MSSM Higgs bosons explaining certain excess LEP events was
(to my knowledge) first suggested in ref.\cite{andre0}, based on preliminary
DELPHI data; this was followed up by refs.\cite{andre}, which speculate about
evidence for three light neutral Higgs bosons lurking in a preliminary version
\cite{prelim}\footnote{This paper estimates the dilution of the total
  significance due to the ``look elsewhere'' effect to be somewhere between 30
  and 60. However, this refers to a {\em two--dimensional} scan of the
  $(m_h,m_A)$ plane, which obviously contains many more independent bins than
  the one--dimensional distribution analyzed in the present paper. Note also
  that the preliminary LEP analysis assigns about equal significance to the
  excesses near 98 and near 115 GeV, showing that it was indeed a {\em
    preliminary} analysis.} of the combined LEP data.  The same preliminary
data are also discussed in \cite{weiglein1}.  Neither of these papers attempts
to explore the phenomenology of scenarios with $m_h \simeq 98$ GeV and
suppressed $ZZh$ coupling. Refs.\cite{kane} discuss several MSSM scenarios
with some neutral Higgs boson(s) below 115 GeV, but again do not explore the
phenomenology; the main focus of these articles is on issues of model building
and finetuning.

The rest of this paper is organized as follows. In Sec.~2 I describe the LEP
Higgs searches in more detail, and show how the excess events can be described
in the MSSM. In Sec.~3 I explore the parameter space that is compatible with
this explanation, and discuss tests of this scenario. Finally, Sec.~4 contains
a brief summary and conclusions.

%%%%%%%%%%%%%%%%%%%%%%%%%%%%%%%%%%%%%%%%%%%%%%%%%%%%%%%%%%%%%%%%%%%%%%%%%%
\section*{2. Searches for the SM Higgs at LEP}

Searches for the single Higgs boson $\varphi$ of the Standard Model at LEP are
based on the process
\be \label{bj}
e^+ e^- \rightarrow Z \varphi .
\ee
The most important decay modes of the Higgs boson are $b \bar b$ and $\tau^+
\tau^-$. The statistically most important final state contains four jets, from
$Z \rightarrow q \bar q$ and $\varphi \rightarrow b \bar b$, but some other
channels are also of interest: $Z \rightarrow \nu \bar \nu, \ \varphi
\rightarrow b \bar b$ leads to large missing energy plus $b-$jets; $Z
\rightarrow \ell^+ \ell^-, \ \varphi \rightarrow b \bar b$ leads to a charged
lepton (electron or muon) pair plus $b-$jets; and $Z \rightarrow \tau^+
\tau^-, \ \varphi \rightarrow b \bar b$ or $Z \rightarrow q \bar q, \ \varphi
\rightarrow \tau^+ \tau^-$ lead to events with a $\tau$ pair and jets.

After combining the data samples of all four LEP experiments and all four
final states, any value of $m_\varphi \leq 114.4$ GeV can be excluded at the
95\% confidence level \cite{leph}. This limit is a little weaker than that
expected for heavy $\varphi$ given the performance of LEP. The reason is that
there is an excess of events near the kinematic end point. This excess was
first announced by ALEPH \cite{aleph}, where it reaches a significance of
about 3 standard deviations; this led to considerable excitement at that
time. Unfortunately the other LEP experiments see little or no excess there;
the combination of LEP data therefore weakened the excess to the level of
about 1.7 standard deviations.

At the same time this combination {\em strengthened} the very mild excess also
reported by ALEPH near 98 GeV. The reason is that all four experiments see a
mild excess here, the most significant being in the L3 data sample where it
reaches about 1.8 standard deviations. Moreover, there is an excess both in
the four--jet channel {\em and} in the sum of the other three channels. All
this is just what one would expect from a true signal at the edge of
statistical detectability.

Unlike the excess near 115 GeV, the accumulation of events near 98 GeV cannot
be interpreted as production of the SM Higgs boson. In the SM the rate for
reaction (\ref{bj}) can be predicted uniquely as a function of the mass of the
Higgs boson; it comes out much too large if $m_\varphi \simeq 98$ GeV. This
excess, if real, therefore calls for physics beyond the Standard Model.

The best motivated extension of the SM has long been its supersymmetric
version, the MSSM \cite{mssm}. Supersymmetrizing the SM stabilizes the mass of
the Higgs boson(s) against large radiative corrections, thereby solving the
technical part of the hierarchy problem \cite{hierarchy}. It also naturally
allows the Grand Unification of all gauge interactions, without the {\it ad
  hoc} introduction of any new particles (beyond those required by
Supersymmetry) \cite{GaugeUni}; and allows for a natural explanation for the
Dark Matter in the Universe whose existence has been proven almost
unambiguously by cosmologists \cite{dm}.

Of greater relevance for the present purpose is that Supersymmetry requires
the existence of at least two Higgs doublets, in order to give masses to both
up--type and down--type quarks, and to cancel gauge anomalies associated with
the fermionic superpartners of the Higgs bosons. In the MSSM one thus
postulates the existence of two Higgs doublets with opposite hypercharges. The
neutral components of both neutral Higgs fields must acquire VEVs to make all
quarks massive. Both VEVs contribute (in quadrature) to the masses of the $W$
and $Z$ bosons; three would--be Goldstone modes get ``eaten'' in the process.
Therefore only five physical degrees of freedom survive in the MSSM Higgs
sector \cite{gh}: two neutral CP--even scalars $h$ and $H$, with $m_h < m_H$;
a CP--odd scalar $A$; and charged Higgs bosons $H^\pm$.\footnote{In the
  presence of CP--violation in the squark sector, the three neutral gauge
  bosons will mix, to form mass eigenstates which are no longer CP
  eigenstates. Given the stringent constraints on electric dipole moments
  \cite{pdg}, which are not easy to satisfy in the presence of large CP--odd
  phases in the supersymmetric Lagrangian, I will ignore this possibility for
  the most part.}

At the tree level the MSSM Higgs sector is completely determined once two
parameters are fixed. The most common choice is the mass $m_A$ of the CP--odd
Higgs boson and the ratio of VEVs $\tan\beta$. The masses $m_{h,H}$ as well as
the mixing angle $\alpha$ in the neutral CP--even Higgs sector are then
derived quantities \cite{gh}. The two angles $\alpha$ and $\beta$ also
determine the couplings of MSSM Higgs bosons to SM particles. In particular,
the couplings relevant to the generalization of the process (\ref{bj}) are
\cite{gh}:
\ben \label{zcoup} \beq
g_{h Z Z} &= g_{\varphi Z Z} \sin(\beta - \alpha); \\
g_{H Z Z} &= g_{\varphi Z Z} \cos(\beta - \alpha).
\eeq \een
The fact that the number of excess events near 98 GeV amounts to about 10\% of
the signal for the SM Higgs with $m_\varphi = 98$ GeV therefore immediately
implies that $\sin^2(\beta-\alpha) \simeq 0.1$. Given that the excess has a
statistical significance of 2.3 standard deviations, this leads to the $1
\sigma$ bounds on this coupling:
\be \label{sambsq}
0.056 \leq \sin^2(\beta-\alpha) \leq 0.144.
\ee
In order to make sure that the excess appears in the right mass range, one
also needs
\be \label{mh}
95 \ {\rm GeV} \leq m_h \leq 101 \ {\rm GeV},
\ee
where the range can be motivated either from the width of the peak in the data
(more precisely: from the width in the dip of one minus the background
confidence level) \cite{leph}, or from the accuracy with which the neutral
Higgs boson masses can be calculated (see below).

Having fixed two parameters in the MSSM Higgs sector fairly accurately, it
would seem that the model is already defined completely. However, the MSSM
Higgs sector is subject to large radiative corrections \cite{radcorr1}. On the
one hand, these are crucial for the viability of the model, since at the
tree--level one has $m_h < M_Z$, in gross conflict with Higgs searches at LEP.
On the other hand, this increases the number of free parameters, and hence
also the possible ranges of the masses of the other Higgs bosons. This will be
explored in more detail in the next Section.

Nevertheless the constraint (\ref{sambsq}) already allows some important
conclusions. To begin with, $\sin^2(\beta-\alpha) \simeq 0.1$ implies
$\cos^2(\beta-\alpha) \simeq 0.9$, i.e. $ZH$ production will occur almost with
SM strength, if it is accessible kinematically. This would allow to {\em
  simultaneously} explain the excess of Higgs--like events around 115 GeV, if
\be \label{mH}
111 \ {\rm GeV} \leq m_H \leq 119 \ {\rm GeV}.
\ee
I have again allowed for a few GeV range to accommodate both the theoretical
and the experimental uncertainties.

In fact, the constraint (\ref{sambsq}) {\em by itself} already implies that
the other MSSM Higgs bosons must not be very heavy. The reason is that for
$m_A^2 \gg M_Z^2$ one observes ``decoupling'' \cite{gh2}: $H, A, H^\pm$ form a
nearly degenerate, heavy $SU(2)$ doublet which does not have Higgs--like
couplings to gauge bosons, i.e. $\cos(\beta-\alpha) \rightarrow 0$ in this
limit, in conflict with (\ref{sambsq}). This argument makes it rather natural
to associate the (small) excess of events near 115 GeV with $ZH$
production.
However, in order to quantify the upper limits on the masses of the other MSSM
Higgs bosons, the relevant parameter space has to be explored in detail. This
is the topic of the next Section.

%%%%%%%%%%%%%%%%%%%%%%%%%%%%%%%%%%%%%%%%%%%%%%%%%%%%%%%%%%%%%%%%%%%%%%%%%%
\section*{3. Testing the scenario}

From the discussion of the previous Section it should be clear that both
groups of Higgs--like excess events observed at LEP can indeed be explained
simultaneously through the production of MSSM Higgs bosons. However, for
quantitative tests it is necessary to explore in detail the region of MSSM
parameter space consistent with this scenario.

To that end a treatment of at least the leading radiative corrections
\cite{radcorr1} to the MSSM Higgs sector is mandatory. This is most easily
done using the effective potential (or, equivalently, Feynman diagrammatic
calculations with vanishing external momentum). Since the entire Higgs
spectrum has to be rather light in this scenario, this should be a good
approximation not only for the light scalar, but also for the other Higgs
bosons. Note also that two--loop correction terms are to date anyway only
known in this limit.

In order to allow an efficient sampling of the parameter space, I only include
corrections from the top--stop and bottom--sbottom sectors, which give the by
far most important contributions. I use the expressions in ref.\cite{dn} for
the pure Yukawa corrections to both the neutral and charged Higgs boson mass
matrices; the mixed electroweak--Yukawa corrections to the neutral Higgs boson
mass matrix (from the ``$D-$term'' contributions to the stop and sbottom
masses) are included using results of ref.\cite{cdl}. Leading higher order QCD
corrections are included by using running quark masses defined at the
appropriate scale in the one--loop effective potential, as described in
refs.\cite{dn,hhh}. The leading SUSY QCD corrections are included through the
gluino--stop and gluino--sbottom ``threshold'' corrections to the top and
bottom mass, respectively; in the latter case the corrections have been
resummed using the formalism of ref.\cite{youichi}. As shown in
ref.\cite{heinemeyer}, this reproduces the full SUSY QCD correction very
accurately. Finally, the leading higher order correction from the top Yukawa
coupling is again included via the running top mass in the effective potential
\cite{hhh,carena}. Recently a full calculation of top Yukawa corrections
became available \cite{pietro}. However, this result is badly behaved in the
$\overline{\rm DR}$ scheme for very large stop mixing parameter; since this
region of the parameter space is relevant here, these corrections are not
included. The calculation performed here should reproduce the neutral MSSM
Higgs masses with an error of about 3 GeV or so \cite{herror,heinemeyer}. This
theoretical uncertainty is reflected in the ranges in (\ref{mh}) and
(\ref{mH}).

Again for reasons of simplicity, I work with a fixed SM top mass, $m_t(m_t) =
171$ GeV (in the $\overline{\rm DR}$ scheme). This corresponds to a pole mass
near 178 GeV, the current central value \cite{d0top}. I also fix $m_b(m_b) =
4.25$ GeV. As final simplification, I have taken the soft breaking parameters
in the stop and sbottom mass matrices to be the same.  This is always true for
the masses of the superpartners of the left--handed squarks, due to $SU(2)$
invariance, but the masses of the $SU(2)$ singlet squarks as well as the two
$A-$parameters could in principle be different.  However, $A_b$ only plays a
minor role, since mixing in the $\tilde b$ sector is either small or, for
large $\tan\beta$, controlled by the supersymmetric mass parameter $\mu$
\cite{mssm}.

Altogether we are thus left with seven free parameters: $\tan\beta, \ m_A, \ 
\mu, \ m_{\tilde t_L}, \ m_{\tilde t_R}, \ A_t, \ m_{\tilde g}$. Note that, in
spite of the simplifying assumptions, we still have five free parameters (not
affecting the Higgs sector at tree--level) to describe four radiative
corrections, to the three independent entries of the mass matrix of CP--even
Higgs bosons and to the mass of the charged Higgs boson. As far as the Higgs
sector is concerned the chosen parameterization should therefore be
sufficiently flexible to cover all possibilities (barring CP--violation). This
seven--dimensional parameter space has been scanned subject to the following
constraints:
\ben \label{cons} \beq
|\mu|, \, m_{\tilde t_R}, \, m_{\tilde t_L}, \, m_{\tilde g} &\leq 2 \ {\rm
  TeV}; \\ 
|\mu|, \, m_{\tilde t_1}, \, m_{\tilde b_1} &\geq 100 \ {\rm GeV}; \\
m_{\tilde g} &\geq 300 \ {\rm GeV}; \\
|A_t|, |\mu|  &\leq 1.5 \left( m_{\tilde t_R} + m_{\tilde t_L} \right); \\
\delta \rho_{\tilde t \tilde b} &\leq 2 \cdot 10^{-3}.
\eeq \een
The first of these constraints is a crude naturalness criterion. Conditions
(\ref{cons}b) ensure that higgsino--like charginos (with mass $\sim |\mu|$) as
well as the lighter physical stop $(\tilde t_1)$ and sbottom $(\tilde b_1)$
states escaped detection at LEP \cite{pdg}. Condition (\ref{cons}c) similarly
guarantees that gluinos were not detected at the Tevatron \cite{pdg}. The
upper bounds (\ref{cons}d) on the parameters determining mixing in the stop
and sbottom sectors have been imposed to avoid situations where $\tilde t$ or
$\tilde b$ fields have non--vanishing VEVs in the absolute minimum of the
scalar potential \cite{CCB}. Finally, (\ref{cons}e) requires the contribution
of stop--sbottom loops to the electroweak $\rho$ parameter \cite{delrho} to be
sufficiently small.  Of course, the constraints (\ref{sambsq}) and (\ref{mh})
also have to be imposed in order to describe the excess near 98 GeV. If the
excess near 115 GeV is to be described by the scenario as well, in addition
the constraint (\ref{mH}) should be imposed. In any case one has to require
$m_H > 111$ GeV in order to make sure that $ZH$ production was not detected at
LEP.\footnote{Recall that I allow a few GeV uncertainty on the calculated
  value of the Higgs boson masses; an experimental bound of 114.4 GeV is thus
  interpreted as a bound on the calculated $m_H$ of 111 GeV.}

\begin{figure}[h!]
\begin{center}
\vspace*{-1.2cm}
\rotatebox{270}{\includegraphics[width=10cm]{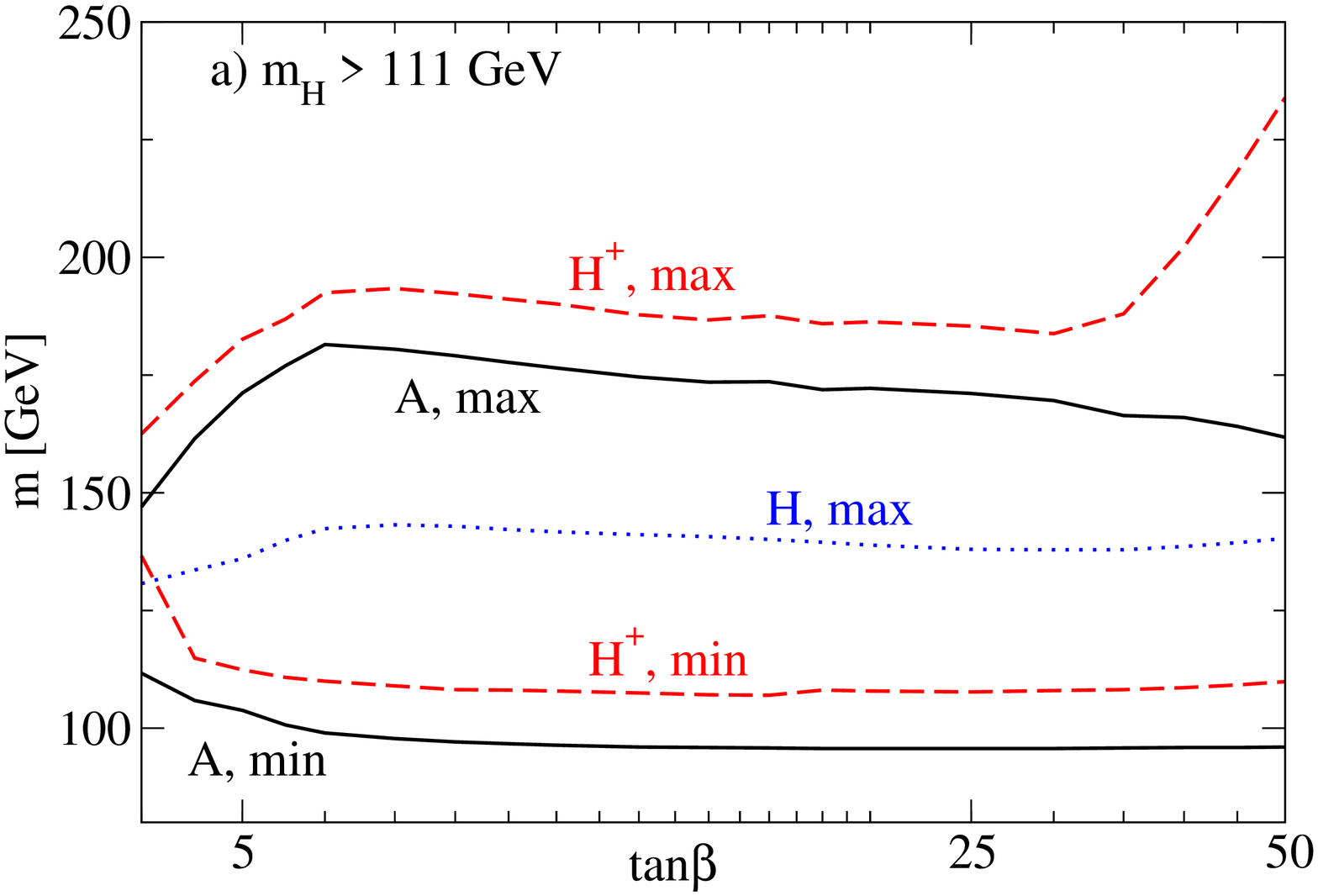}} \\
\vspace*{-.7cm}
\rotatebox{270}{\includegraphics[width=10cm]{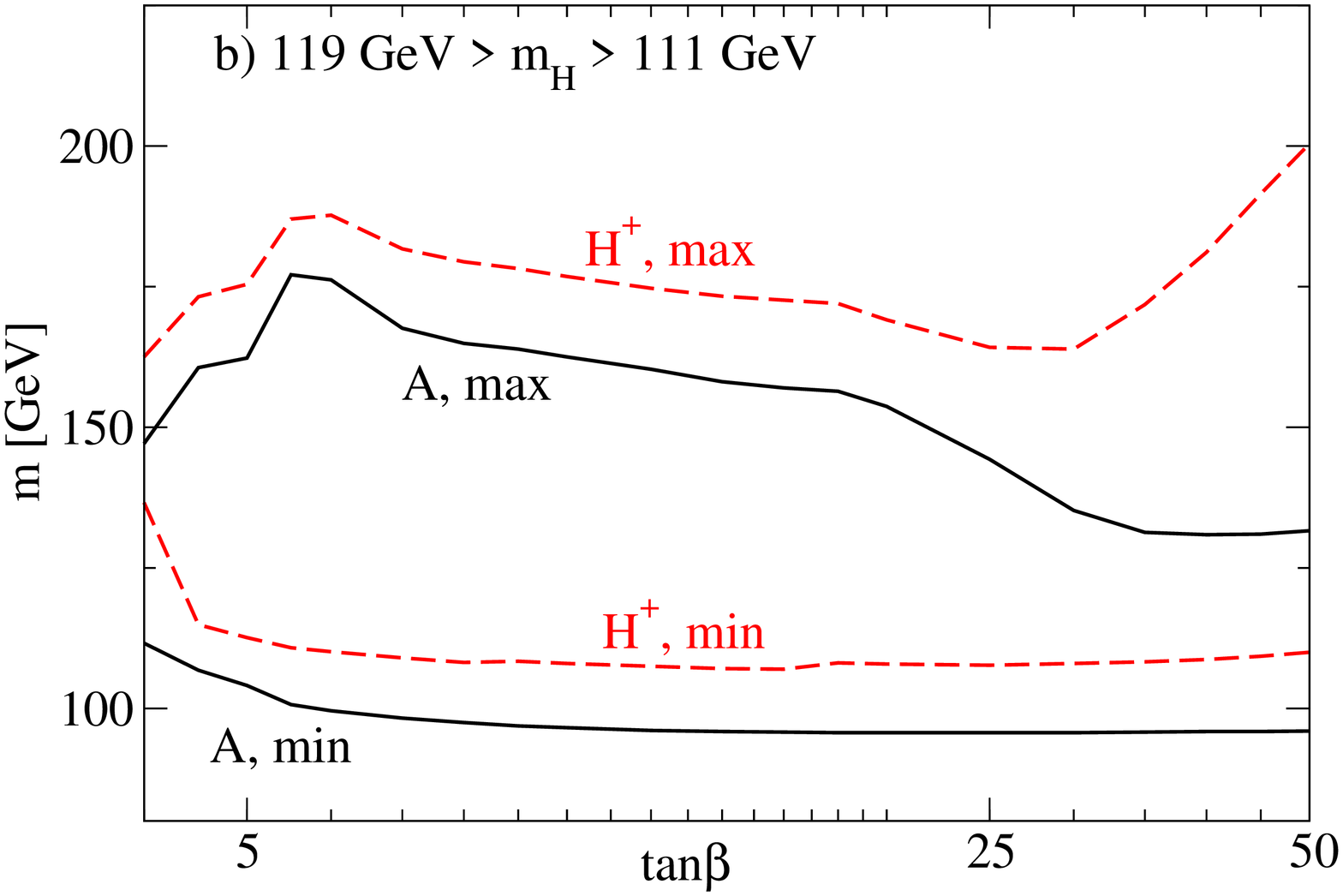}} \\
%\vspace*{7cm}
\caption{%
  Minimal and maximal values of the heavier MSSM Higgs boson masses consistent
  with the constraints (\ref{sambsq}), (\ref{mh}) and (\ref{cons}), as
  required to reproduce the excess of Higgs--like events near 98 GeV. In the
  lower frame in addition the constraint (\ref{mH}) has been imposed, as
  required to {\em also} reproduce the excess near 115 GeV.}
\label{fig1}
\end{center}
\end{figure}

The maximal and minimal allowed masses of the heavier MSSM Higgs bosons that
are compatible with all these constraints are shown in Figs.~1, without (1a)
and with (1b) including the requirement (\ref{mH}). These results have been
obtained by randomly generating several million combinations of $m_A, \ \mu, \ 
m_{\tilde t_L}, \ m_{\tilde t_R}, \ A_t, \ m_{\tilde g}$ for each value of
$\tan\beta$, with special emphasis on the regions of parameter space where one
of the shown masses takes its extremal value. 

We see that the excess near 98 GeV cannot be reproduced for very small values
of $\tan\beta$. If $\tan\beta < 3.7$, $m_h$ can only be sufficiently large if
$m_A$ is also quite large (and stop squarks are quite heavy). This is in
conflict with the requirement (\ref{sambsq}) of a suppressed $ZZh$
coupling. However, the scenario still allows a wide range for $\tan\beta$,
including quite large values. I have not explored the parameter space for
$\tan\beta > 50$, since scenarios with very large $\tan\beta$ and small $m_A$
are strongly constrained by the non--observation of $B_s \rightarrow \mu^+
\mu^-$ decays \cite{bs}.

As anticipated, the scenario is only viable for moderate masses of the heavier
Higgs bosons. Starting from the minimal allowed value of $\tan\beta$, the
upper bounds first increase, since the allowed parameter space opens up. $m_A$
reaches its absolute maximum of slightly more than 180 GeV at $\tan\beta
\simeq 6$. The bound on $m_A$ then decreases again slowly. The reason is that
the cross--over from almost vanishing to essentially maximal $ZZh$ coupling
becomes faster as $\tan\beta$ increases, i.e. the ``decoupling regime'' is
approached more quickly for larger $\tan\beta$ \cite{gh2}. The upper bound on
the mass of the charged Higgs boson first tracks that on $m_A$. However, for
$\tan\beta > 30$ the radiative corrections to $m_{H^\pm}$ become sizable,
eventually allowing $m_{H^\pm}$ well above 200 GeV. Note also that the lower
limits on $m_A$ and $m_{H^\pm}$ are so high that searches for $hA$ and $H^+
H^-$ production at LEP do not constrain this scenario any further; due to the
$P-$wave suppression of these cross sections, these searches were only
sensitive to Higgs masses below $\sim 90$ GeV \cite{pdg}.

Finally, Fig.~1a also shows that the excess near 98 GeV can only be reproduced
if $m_H < 145$ GeV. (The lower limit on $m_H$ is just given by the
experimental bound of 111 GeV discussed above.) This indicates that explaining
the (small) excess of events near 115 GeV through $ZH$ production is quite
natural in this framework. Fig.~1b shows how the limits on $m_A$ and
$m_{H^\pm}$ are tightened when this is done, i.e. if in addition the
constraint (\ref{mH}) is imposed. We see that the lower limits on these masses
remain essentially unchanged. However, the upper bound on $m_A$ is reduced
significantly, especially at large $\tan\beta$. The reason is that the
tree--level $A-H$ mass splitting decreases with increasing $\tan\beta$; the
increased importance of $b-\tilde b$ loops can compensate this only partly.
Note, however, that charged Higgs boson masses somewhat above the mass of the
top quark can still be realized both for $\tan\beta \simeq 6$ and for large
$\tan\beta$.

The results of Figs.~1 allow us to discuss tests of this
scenario. The constraint (\ref{sambsq}) immediately determines the $WWh$ and
$ZAh$ couplings, which are proportional to $\sin(\beta-\alpha)$ and
$\cos(\beta-\alpha)$, respectively \cite{gh}. This means that $h$ production
at the Tevatron, which relies mostly on $Wh$ and $Zh$ production, will be
impossible to detect \cite{tevh}. The large $ZAh$ coupling together with the
upper bound on $m_A$ means that any (linear) $e^+e^-$ collider operating at
$\sqrt{s} \gsim 300$ GeV should see a strong signal for $Ah$ production; if
the beam energy is raised to $\sim 500$ GeV, $H^+H^-$ production should also
be visible over the entire allowed parameter range in this scenario \cite{lc}.
However, presently there is no funding for the construction of such a device.

\begin{figure}[h!]
\begin{center}
\vspace*{-,7cm}
\rotatebox{270}{\includegraphics[width=10.cm]{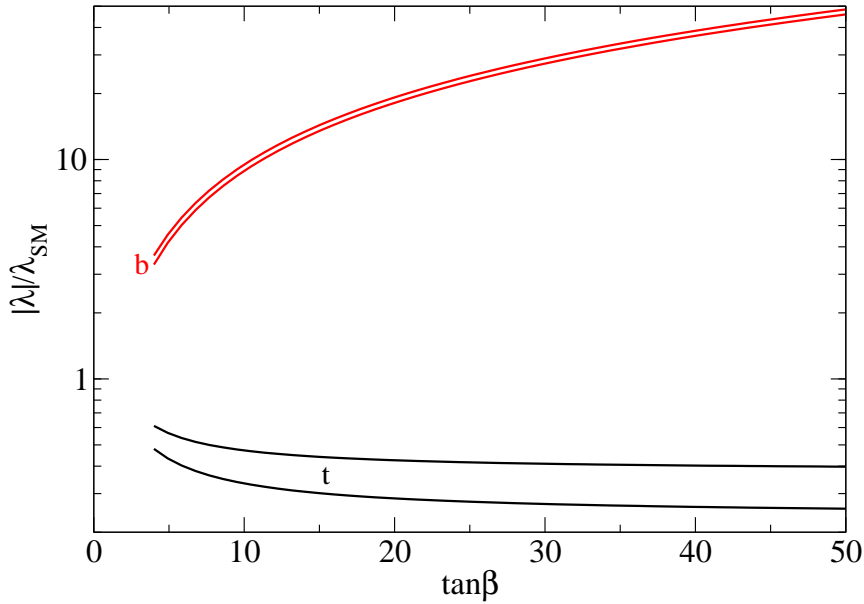}} \\
\caption{%
  The couplings of the light CP--even MSSM Higgs boson $h$ to heavy quarks as
  a function of $\tan\beta$ in units of the corresponding SM coupling, once
  the constraint (\ref{sambsq}) has been imposed. The upper and lower curves
  within a pair should be interpreted as defining the $1 \sigma$ bands for
  these couplings, since they have been obtained by saturating the limits in
  (\ref{sambsq}).}
\label{fig2}
\end{center}
\end{figure}

At least within the next decade or so, the best chance to test this scenario
will therefore be at the LHC. Unfortunately the suppressed $WWh$ and $ZZh$
couplings mean suppressed $h$ production rates through $WW$ and $ZZ$ fusion
\cite{dieter}. The strengths of other potential signals for $h$ production at
the LHC also depend on its couplings to heavy quarks. These depend on
$\tan\beta$ even if the constraint (\ref{sambsq}) is imposed, as shown in
Fig.~2. The increased coupling strength to $b \bar b$ pairs and the reduced
$WWh$ coupling imply a greatly reduced branching ratio for $h \rightarrow
\gamma \gamma$ decays, which mostly proceeds through loops of $W$ bosons
\cite{gh2} and gives the best signal for inclusive $h$ production at the LHC
\cite{lhch}.  At the same time, the suppressed coupling to top quarks means
that the inclusive $h$ production cross section through gluon fusion will also
be suppressed at small and moderate values of $\tan\beta$. The same
suppression applies to the cross section for $t \bar t h$ production.

On the other hand, the cross section for $h b \bar b$ production will be
enhanced at large $\tan\beta$. The LHC will therefore be able to probe a
significant fraction of the allowed parameter space through $(A,h) b \bar b$
production followed by $A,h \rightarrow \mu^+ \mu^-$ decays
\cite{mumu,lhch}. However, Figs.~1 show that, unlike in the standard scenarios
usually considered, the $h$ and $A$ masses can still differ by several tens of
GeV even at large $\tan\beta$, so that the two signals will lead to two
separate peaks, which reduces the significance in any one peak. It is
therefore presently not clear how small values of $\tan\beta$ can be probed
in this scenario through this channel at the LHC. It seems certain, however,
that it will not be able to cover the whole allowed parameter space.

This leaves the charged Higgs boson. We see that over most of the parameter
space, $t \rightarrow H^+ b$ decays are possible. This opens the possibility
to test this scenario even at the Tevatron \cite{tevch}. However, even if
these decays are allowed, their branching ratio becomes quite small for
$\tan\beta \sim \sqrt{m_t(m_t)/m_b(m_t)} \simeq 7$. Besides, we saw in Fig.~1b
that $m_{H^+} > m_t-m_b$ remains possible even if both sets of excess events
at LEP are to be explained by MSSM Higgs production. At the LHC the $H^+ \bar
t b$ production channel can be detectable even if $m_{H^\pm} > m_t - m_b$, but
again this only works at relatively large $\tan\beta$ \cite{lhch,moretti}. One
thus has to conclude that discovery of a not--SM--like Higgs boson at the LHC
does not seem to be guaranteed in this scenario.

Of course, the heavy neutral scalar $H$ should be quite easy to discover at
the LHC even in the less constrained scenario depicted in Fig.~1a, e.g.
through $WW/ZZ$ fusion \cite{dieter}. This will also tell us whether the
excess of events near 115 GeV is indeed due to Higgs boson production or
merely a background fluctuation. However, the cross section will differ from
the corresponding SM cross section only by $\sim 10$\%, well below the
foreseen accuracy with which it can be measured at the LHC even at high
luminosity \cite{dietercoup}. Failing to find a signal for $H$ production
below $\sim 145$ GeV will therefore exclude this scenario, but finding such a
signal will not help to distinguish it from the SM (nor from a more generic
MSSM scenario with SM--like $h$).

\begin{figure}[h!]
\begin{center}
\vspace*{-,7cm}
\hspace*{-.7cm}
\rotatebox{270}{\includegraphics*[width=9cm]{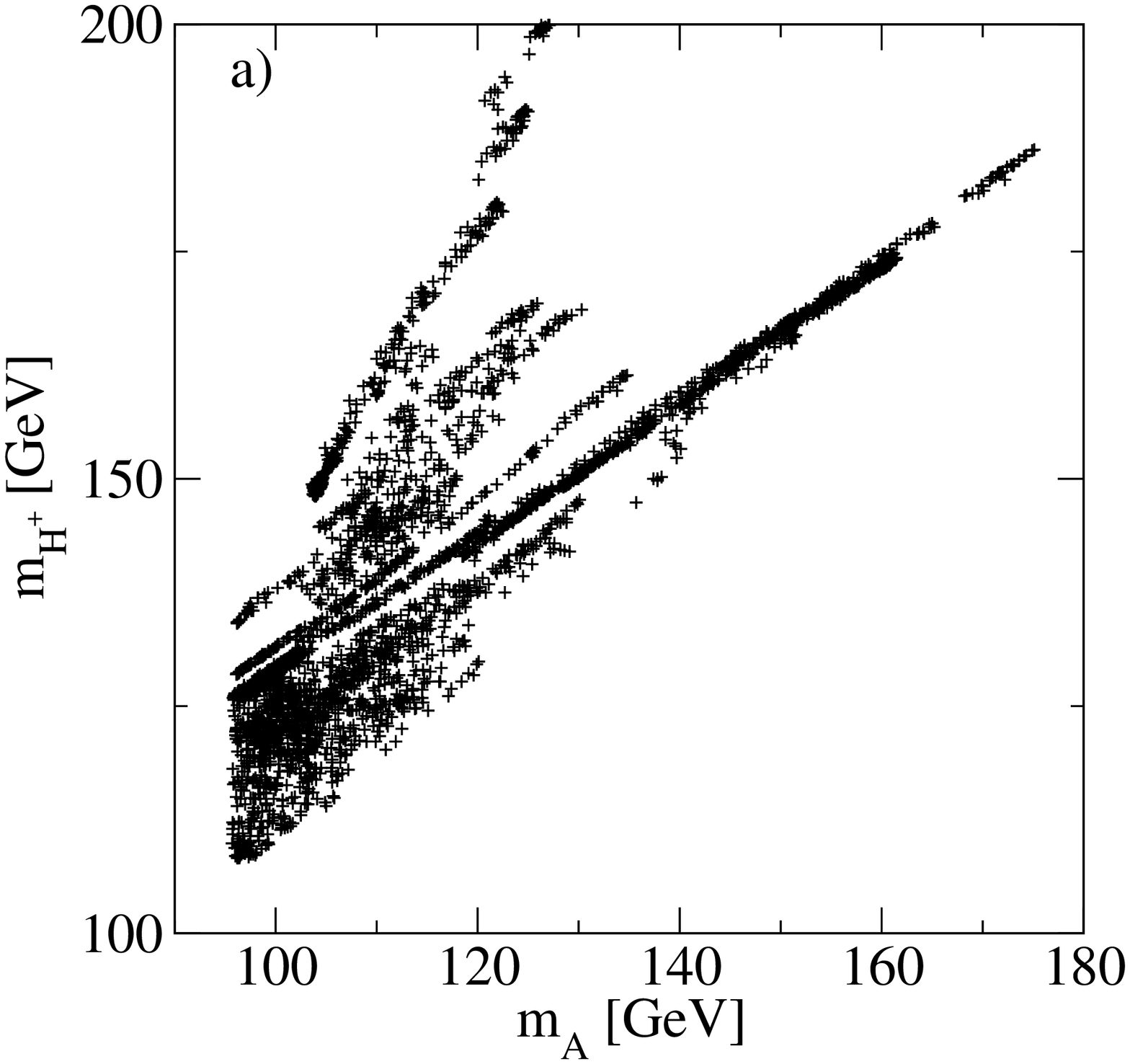}} \hspace*{-2.5cm}
\rotatebox{270}{\includegraphics*[width=9cm]{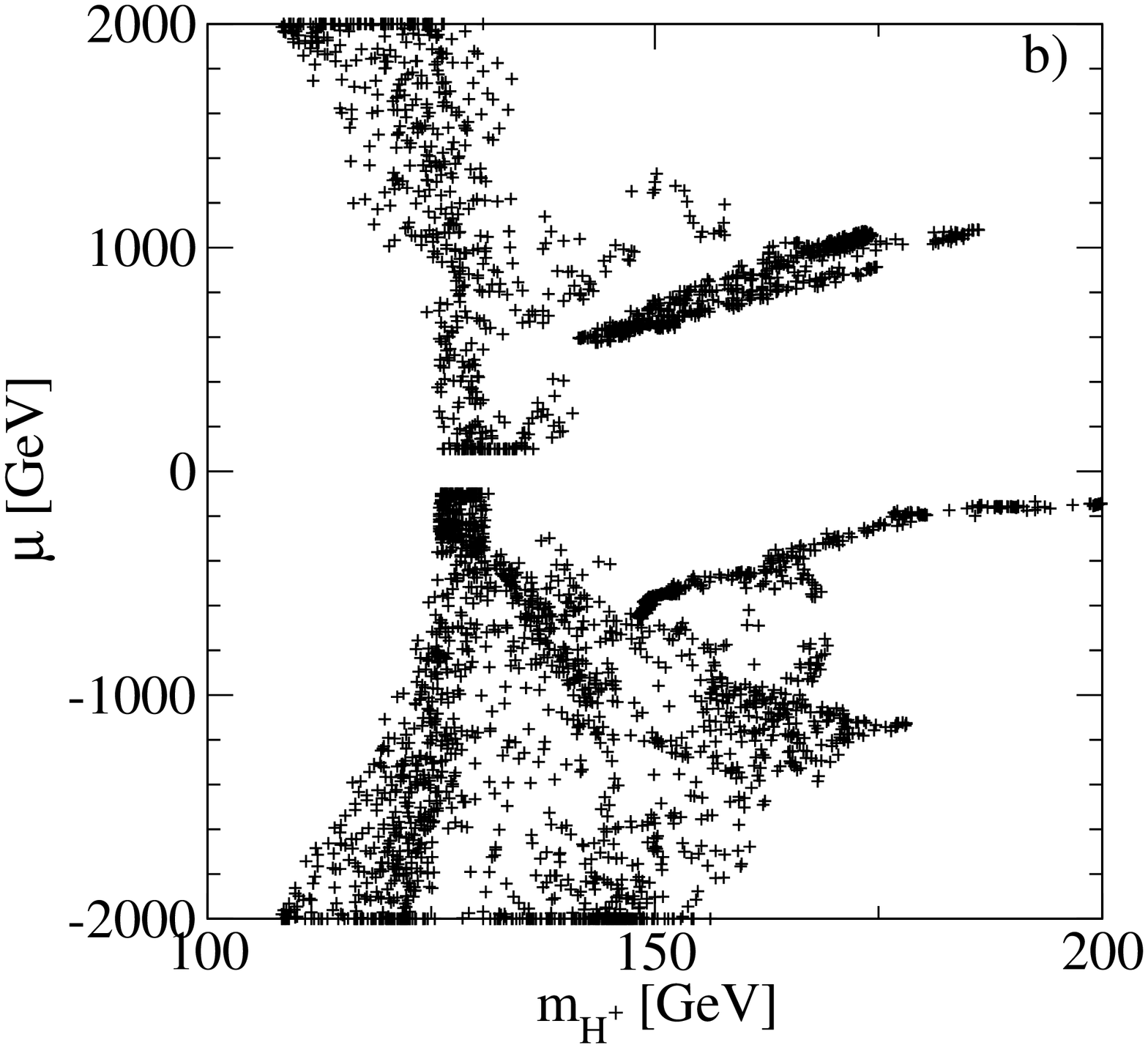}} \\
\hspace*{-.7cm}
\rotatebox{270}{\includegraphics*[width=9cm]{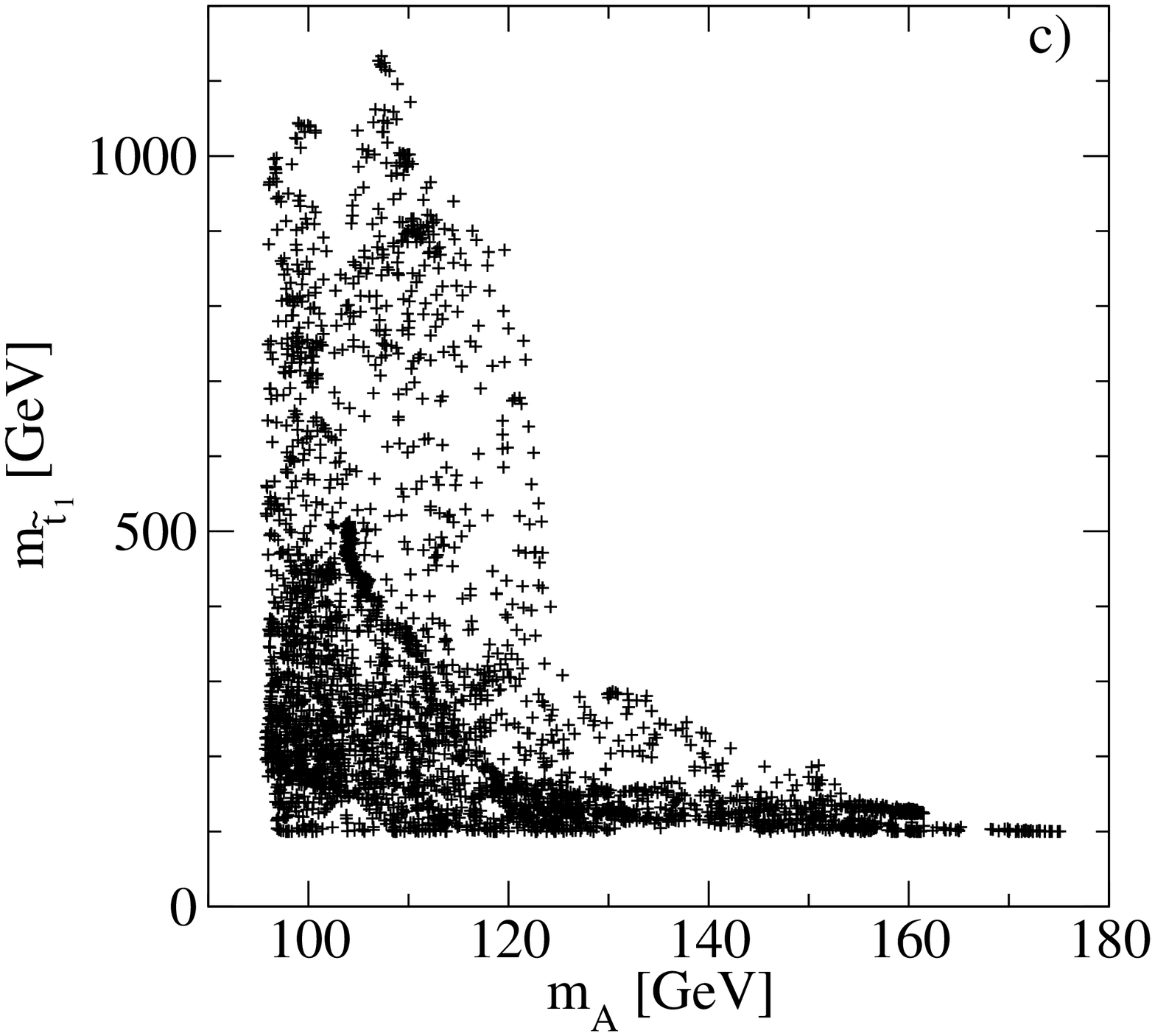}} \hspace*{-2.5cm}
\rotatebox{270}{\includegraphics*[width=9cm]{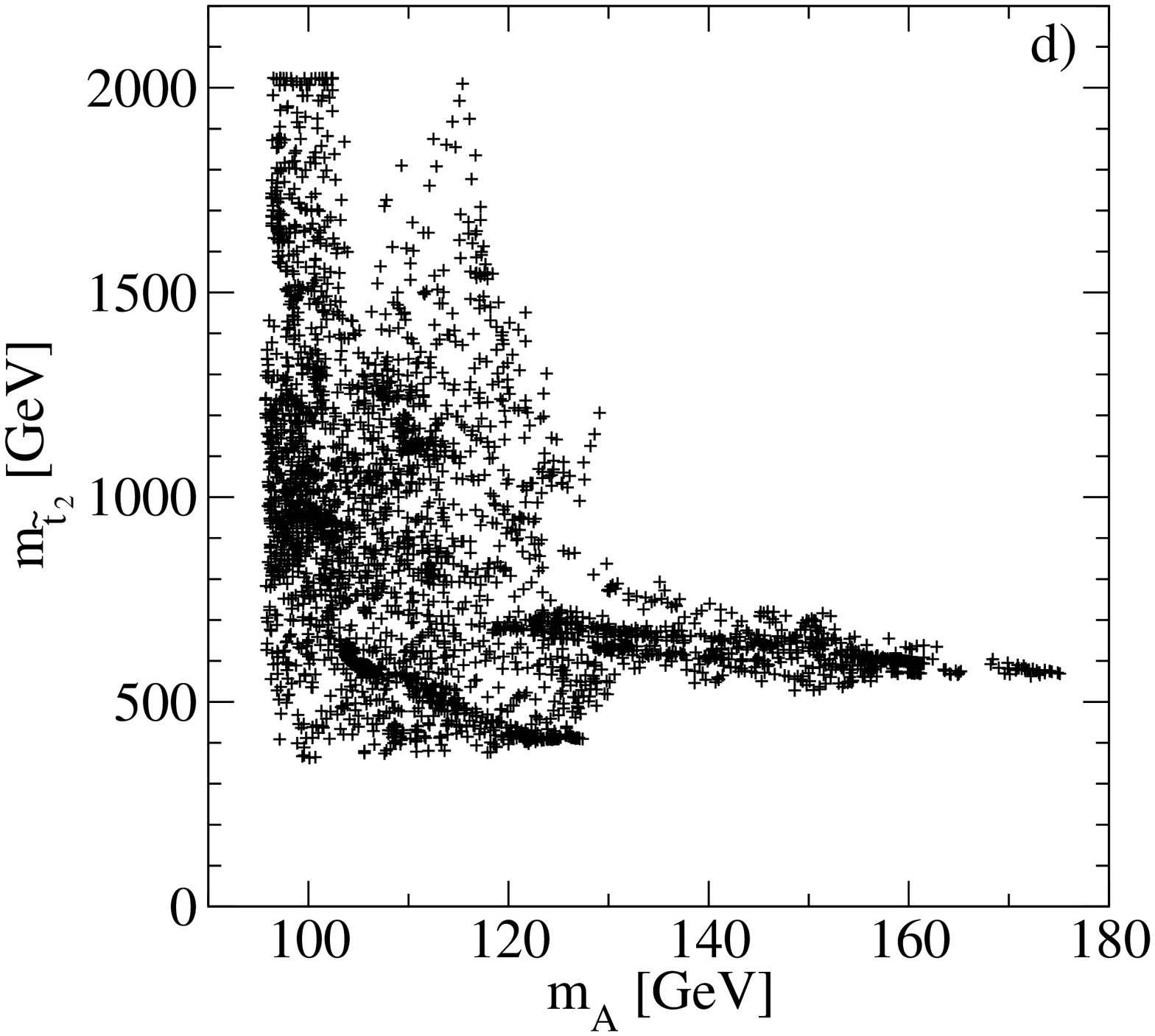}} \\
\caption{%
Allowed region in the $(m_A, m_{H^\pm})$ plane (a), in the $(m_{H^\pm},\mu)$
plane (b), in the $(m_A, m_{\tilde t_1})$ plane (c) and in the $(m_A,
m_{\tilde t_2})$ plane (d), after the constraints (\ref{sambsq})--(\ref{cons})
have been imposed. }
\label{fig3}
\end{center}
\end{figure}

The constraints (\ref{sambsq})--(\ref{mH}) also lead to correlations between
the free parameters, including the parameters of the stop sector. Some of
these are depicted in Figs.~3a--d. I should stress that the density of points
in these figures does not carry meaningful information; only the extent of the
regions of parameter space populated by (some) points is significant.
Moreover, the allowed ranges for the physical stop masses could presumably be
extended somewhat by including additional loop corrections, e.g. those due to
electroweak interactions, which have been omitted here. Due to the essentially
logarithmic dependence of the leading loop corrections on the stop masses,
subleading corrections could modify the allowed stop masses significantly even
if they have only minor impact on the masses of the Higgs boson.  The
qualitative trends shown in Fig.~3 should nevertheless survive in a more
complete calculation.

Fig.~3a shows that the strict correlation between the tree--level $A$ and
$H^\pm$ masses, $m^2_{H^\pm,\,{\rm tree}} = m^2_{A,\,{\rm tree}} + M_W^2$,
can get loosened significantly once loop corrections are included. In the
allowed parameter space this tree--level relation still holds to good
approximation for $m_A > 140$ GeV. We saw from Fig.~1b that such heavy CP--odd
Higgs bosons are only possible in the present framework if $\tan\beta$ is not
large. On the other hand, the corrections to $m_{H^\pm}$ become maximal at
large $\tan\beta$, as also shown in Fig.~1a. In particular, the branch of
points in Fig.~3a with $m_A$ near 125 GeV, which extends to $m_{H^\pm}
\simeq 200$ GeV, corresponds to solutions with $\tan\beta \geq 40$. 

The correlation between the mass of the charged Higgs boson and the higgsino
mass parameter $\mu$ is explored in Fig.~3b. We see that viable scenarios with
$|\mu|$ near the experimental lower bound of $\sim 100$ GeV can easily be
found. On the other hand, significant negative corrections to $m_{H^\pm}$ are
possible only for large values of $\mu$. In contrast, the two branches in
Fig.~3a leading to relatively heavy charged Higgs bosons correspond to more
moderate values of $|\mu|$: the branch where $m_A$ is near its maximum
corresponds to the accumulation of points near $\mu = 1$ TeV in Fig.~3b, while
the absolute maximum of $m_{H^\pm}$ is reached for $\mu \simeq -200$ GeV.
This latter part of the parameter space can therefore also be explored through
the production of higgsino--like neutralinos and charginos at an $e^+e^-$
collider operating at $\sqrt{s} \gsim 500$ GeV \cite{lc}.

The correlations between the mass of the CP--odd Higgs boson and the masses of
the two physical stop states are shown in Figs.~3c,d. Fig.~3c show that a mild
upper bound of about 1.2 TeV on the mass of $\tilde t_1$ can be derived in
this scenario. However, this will be difficult to test even at the LHC in the
presence of other supersymmetric backgrounds; besides, as argued above, this
bound might be modified significantly once purely electroweak corrections to
the Higgs sector are included. Of more interest is the observation that large
CP--odd Higgs boson masses, $m_A \gsim 130$ GeV, are only possible in this
scenario if $m_{\tilde t_1} \lsim 300$ GeV and $m_{\tilde t_2} \lsim 800$ GeV.
In particular, the accumulation of points near the lower bound on $m_{\tilde
  t_1}$ indicates that even Tevatron searches for $\tilde t_1$ production
\cite{tevstop} should be able to explore some significant fraction of the
parameter space of this scenario. Note also that the {\em heavier} stop
eigenstate might be as light as 350 GeV in this scenario. One does not need
large loop corrections to lift $m_h$ from its tree--level upper bound of $M_Z$
to values near 98 GeV. 

\section*{4. Summary and Conclusions}

In this paper I have shown that the $\sim 2.3 \sigma$ excess of Higgs--like
events observed by all four LEP experiments near a Higgs boson mass of 98 GeV
can easily be described by $Zh$ production in the MSSM, where $h$ is the
lighter of the two CP--even Higgs bosons in this model. This explanation
requires relatively small masses for the other Higgs bosons: $m_H \lsim 145$
GeV, $m_A \lsim 180$ GeV and $m_{H^\pm} \lsim 230$ GeV. It is then tempting to
explain the $\sim 1.7 \sigma$ excess of Higgs--like events at LEP near a Higgs
boson mass of 115 GeV through the production of the heavier CP--even state
$H$. The total excess then slightly exceeds the level of 3 standard
deviations. This is of course far from compelling, but does appear intriguing.

A Standard Model like Higgs boson with mass near 115 GeV would be easy to
discover at the LHC; some evidence for it might even be found at the Tevatron.
However, this by itself does not tell us anything about the excess of events
near 98 GeV. In order to fully test this scenario, one will have to detect one
of the Higgs bosons that can {\em not} be confused with the single Higgs boson
of the SM. The best chance for the Tevatron would be $t \rightarrow H^+ b$
decays; at the LHC the charged Higgs boson can also be discovered through
non--resonant $\bar t b H^+ $ production. However, these channels will
probably not cover the entire allowed parameter space. Similarly, over much of
the parameter space the LHC should see associate $b \bar b (h,A)$ production
where the Higgs boson decays into a $\mu^+ \mu^-$ pair, but again this will
not cover the entire parameter space. A decisive test may only be possible at
a future (linear) $e^+ e^-$ collider, which would easily be able to detect $h
A$ and $H^+ H^-$ pair production over the entire allowed parameter space.

We also saw that this scenario is consistent with rather small stop and
higgsino masses; the former are even required if the mass of the CP--odd Higgs
boson is near its upper bound. The search for $\tilde t_1$ pair production at
the Tevatron will therefore cover part, but again not all, of the allowed
parameter space. The fact that this scenario can tolerate a rather light
sparticle spectrum also means that it is technically more natural than the
standard choice $m_h > 114$ GeV, which requires large radiative corrections,
and hence large stop masses, which in turn leads to some amount of finetuning
in the Higgs sector to keep $M_Z$ at its measured value.

In fact, radiative $b \rightarrow s \gamma$ decays indicate that some
sparticles should be relatively light in this scenario, since a charged Higgs
boson with mass below 200 GeV would tend to give too large a branching ratio
for this decay \cite{bsgh}, unless there are compensating sparticle loop
contributions \cite{bsgs}. However, in order to make this statement
quantitative, one would have to specify the flavor structure of the soft
breaking masses, which is very model dependent.

In this paper I have taken all relevant parameters directly at the weak scale,
and have treated them as independent quantities. In constrained scenarios
with a simple superparticle spectrum at some high energy scale a small mass
for the CP--odd Higgs boson can only be realized for large values of the
ratio of VEVs $\tan\beta$. This is true both for mSUGRA, where all scalars
receive the same soft breaking mass at a scale near $10^{16}$ GeV \cite{dn0},
and for the minimal model with gauge--mediated supersymmetry breaking
\cite{gmsb}. This can be problematic, since, as mentioned earlier, models with
small $m_A$ and large $\tan\beta$ are strongly constrained by $B_s \rightarrow
\mu^+ \mu^-$ decays \cite{bs}. It is presently not clear whether the scenario
proposed in this paper can be realized in these constrained models.

In this analysis all parameters have been assumed to be real. In the presence
of nontrivial complex phases, CP is violated, and all three neutral Higgs
bosons will mix at the one--loop level \cite{cph,cdl}. In this case the excess
events near 98 GeV or those near 115 GeV might be due to the production of two
nearly degenerate Higgs bosons. This would (almost) fix the masses of all
three neutral Higgs bosons, making such a scenario even more constrained than
the case discussed here. If all mass splittings are large, the mixing between
CP--even and CP--odd states tends to be suppressed, leaving Higgs
phenomenology essentially unaltered.

Of course, the excess of Higgs--like events found at LEP can also be explained
in a non--supersymmetric model with two Higgs doublets \cite{hunter}. However,
in that case no predictions for the masses and couplings of the CP--odd Higgs
boson or the charged Higgs boson can be made, since the Higgs potential
contains many more free parameters than that of the MSSM. Finding these
particles within the limits indicated in Fig.~1 would therefore be strong
indirect evidence in favor of supersymmetry.

To conclude, LEP experiments may already have found the first indication for
the production of neutral MSSM Higgs bosons. This hypothesis would be
strengthened by observing a light charged Higgs boson at the Tevatron and/or
the LHC, or by observing neutral Higgs bosons at LHC that do not resemble the
single Higgs boson of the Standard Model. It would be refuted if the LHC does
not find an SM--like Higgs boson with mass below $\sim 145$ GeV, which
however, would also exclude much of the general MSSM parameter space. Decisive
tests of this scenario may only be possible at future linear $e^+e^-$
colliders.

\end{document}